\documentclass[accepted]{uai2025} 
                        
\usepackage[american]{babel}

\usepackage{natbib} 
\usepackage{mathtools} 
\usepackage{booktabs} 
\usepackage{tikz} 

\usepackage{multirow}

\usepackage{amsmath}
\usepackage{amssymb}
\usepackage{amsthm}

\usepackage[capitalize,noabbrev]{cleveref}

\usepackage{algorithm}
\usepackage{algorithmic}

\newcommand{\Lc}{\mathcal{L}}
\newcommand{\Dc}{\mathcal{D}}
\newcommand{\Eb}{\mathbb{E}}
\newcommand{\n}{\mathbf{n}}

\theoremstyle{plain}

\theoremstyle{definition}

\theoremstyle{remark}

\title{Improved Algorithms for Differentially Private Language Model Alignment}

\author[1]{Keyu Chen}
\author[1]{Hao Tang}
\author[1]{Qinglin Liu}
\author[1]{Yizhao Xu}

\affil[1]{%
    Peking University
}
  
\begin{document}
\maketitle

\begin{abstract}
Language model alignment is crucial for ensuring that large language models (LLMs) align with human preferences, yet it often involves sensitive user data, raising significant privacy concerns. While prior work has integrated differential privacy (DP) with alignment techniques, their performance remains limited. In this paper, we propose novel algorithms for privacy-preserving alignment, and rigorously analyzing their effectiveness across varying privacy budgets and models. Our framework can be specialized to two celebrated alignment techniques, namely direct preference optimization (DPO) and reinforcement learning from human feedback (RLHF). Through systematic experiments on large-scale language models, we demonstrate that our approach achieves state-of-the-art performance. Notably, one of our algorithms, namely DP-AdamW, combined with DPO surpasses existing methods, improving alignment quality by up to 15\% under moderate privacy budgets ($\varepsilon$=2–5). We further investigate the interplay between privacy guarantees, alignment efficacy, and computational demands, providing practical guidelines for optimizing these trade-offs. 

\end{abstract}    

\section{Introduction}

Large language models (LLMs) have demonstrated remarkable capabilities across various tasks, yet ensuring their outputs align with human preferences and values remains a critical challenge~\citep{patil2024review}. Recent advances in alignment techniques, such as Direct Preference Optimization (DPO)~\citep{saeidi2024insights} and Proximal Policy Optimization (PPO)~\citep{li2023policy}, have shown promising results in adapting these models to better reflect human intent~\citep{xu2024dpo}. However, these alignment methods typically require access to extensive human feedback data, raising significant privacy concerns about the potential exposure of sensitive information contained in training examples. For example, ~\citep{carlini2022quantifying} demonstrated that large language models can memorize and reproduce verbatim sequences from their training data, including personal information such as email addresses and phone numbers. These privacy risks are particularly acute in alignment scenarios, where training data often includes personal preferences, opinions, and potentially sensitive user interactions that could be used to identify individuals or reveal private information.

To mitigate the privacy issues, one of the promising approach it to leverage the notion of differential privacy (DP)~\citep{dwork2014algorithmic}. Roughly speaking, DP requires that when a training data alternates, the output, or the trained model does not significantly change. Despite the potential of DP to protect training data, 
its application to language model alignment remains under explored. Existing works~\citep{behnia2022ew,wu2023privately,charles2024fine} rely on DP-SGD~\citep{abadi2016deep}, a classic differentially private algorithm for deep learning. However, stochastic gradient descent (SGD) may not be the best choice for training language models, and it typically uses ADAM~\citep{kingma2014adam} or ADAMW~\citep{loshchilov2017decoupled}. 

Moreover, there are many different techniques developed for the alignment of language models. 
Yet, existing approaches either focus solely on alignment quality without privacy considerations~\citep{xiong2024iterative}, or address privacy in standard fine-tuning scenarios without considering the unique requirements of alignment tasks~\citep{mattern2022limits, hu2023differentially}.

In this paper, we aim to address the aforementioned challenge by investigating the following question: {
\begin{center}\it Can we achieve high performance of language model alignment while providing rigorous privacy guarantees for the training data?
\end{center}} 
We provide affirmative answer to the question. Specifically, we unify existing alignment techniques and provide a differentially private algorithm for the alignment. Our experiments show that the proposed algorithm is better than DP-SGD based private alignment techniques. 
We summarize our contributions in the following.

1. We propose a unified framework for privacy-preserving language model alignment, that consists of a sequence of losses minimization. This unified framework includes current commonly adopted alignment techniques, namely reinforcement learning from human feedback (RLHF) and DPO, as special cases. 

2. We develop a new private optimizer, namely DP-ADAMW, which incorporates the decoupled weight decay into DP-ADAM. More importantly, by applying the private optimizer to the aforementioned unified alignment framework, we obtain our novel differentially private language model alignment algorithm. 

3. We conduct extensive experiments on LLAMA-8B and GPT-2, DeepSeek-LLM-7B-Chat. Specifically, we examine our proposed algorithm in three different dimension. First, we compare our algorithm with existing methods that uses DP-SGD, which shows that our method and its specialization to DP-ADAM achieves better performance for privately aligned language model. Second, we compare different models with our proposed algorithm, which shows the generalization of our algorithm. Finally, we intensively examine the effects of different privacy budget on the performance of the fine-tuned language model.


4. Through analyzing our experiments, we establish practical guidelines for selecting privacy budgets and optimization strategies, offering concrete recommendations for balancing privacy protection with alignment quality in different deployment scenarios. Our results demonstrate that effective model alignment can be achieved while maintaining strong privacy guarantees, though careful consideration must be given to the choice of optimization method and privacy budget. We identify DP-ADAMW and DPO as particularly promising approaches, especially compared to existing approaches using DP-SGD and RLHF~\citep{wu2023privately}.

\section{Related Work}

\subsection{Reinforcement Learning from Human Feedback}
Reinforcement learning from human feedback (RLHF) has emerged as a transformative approach in language model fine-tuning. Unlike conventional methods relying on large labeled datasets, RLHF harnesses human feedback to generate reward signals that guide model optimization, enabling more desirable outputs in complex, open-ended tasks. The seminal work by \citet{christiano2017deep} established the foundational framework, introducing human feedback for reward modeling coupled with Proximal Policy Optimization (PPO) \citep{schulman2017proximal} for model training.

Initial applications of RLHF in natural language processing focused on specific tasks such as stylistic text continuation and summarization \citep{ziegler2019fine,stiennon2022learning,wu2021recursively}, as well as machine translation \citep{nguyen2017reinforcement,kreutzer2018can}. The field subsequently evolved toward developing AI assistants aligned with human values across diverse instruction-following tasks \citep{ouyang2022training,bai2022training,touvron2023llama}.

\subsection{Differential Privacy in Language Models}
The memorization capabilities of language models \citep{carlini2022quantifying} have led to various privacy vulnerabilities, including training data extraction and membership inference attacks \citep{carlini2019secret,carlini2021extracting,elmahdy2022privacy,mattern2023membership}. To address these security concerns, differentially private (DP) fine-tuning has emerged as a promising defensive strategy for privacy preservation.

Recent works have demonstrated the efficacy of DP-SGD \citep{abadi2016deep} in fine-tuning language models \citep{li2021large}. These studies show that through careful hyperparameter selection and parameter-efficient techniques such as LoRA \citep{hu2021lora}, it is possible to develop language models that maintain competitive performance while providing robust privacy guarantees. A parallel research direction explores private synthetic text generation through DP fine-tuning of pre-trained models \citep{mattern2022differentially,yue2022synthetic}, producing synthetic texts that ensure privacy while preserving utility.

\subsection{Differential Privacy in Reinforcement Learning}
Research at the intersection of differential privacy and reinforcement learning dates back to the foundational work of \citet{balle2016differentially}. Subsequent studies have explored various aspects of this integration, with \citet{wang2019privacy} focusing on Q-learning and introducing noise to value function approximation to achieve differential privacy guarantees.

The field has continued to evolve with specialized approaches for different scenarios. \citet{ma2019differentially} address the specific case of Markov Decision Processes (MDPs) with linear function approximations, developing methods to ensure joint differential privacy (JDP). More recently, \citet{qiao2024offline} have extended privacy guarantees to offline datasets, particularly focusing on offline RL algorithms such as Adaptive Policy Value Iteration (APVI) \citep{yin2021towards}.

Despite these advances in privacy-preserving language models and reinforcement learning, there remains a significant gap in ensuring differential privacy for model alignment. To the best of our knowledge, our work represents the first attempt to address this crucial challenge.

\section{Preliminaries}
\subsection{Language Model Alignment}
Language model alignment refers to the process of adapting pretrained language models to better reflect human preferences and values. The alignment process typically begins with Supervised Fine-Tuning (SFT), followed by either RLHF or DPO. In the following, we briefly introduce these two pipelines for completeness. Note that we denote $\pi_{\theta}$ the language model, where $\theta$ is the parameter.

{\bf Dataset.} A typical dataset $\mathcal{D}$ for preference-based alignment consists of triplets $(x, y^+, y^-)$, where $x$ is a prompt, $y^+$ is the preferred response, and $y^-$ is the dispreferred response. These labeled pairs provide the foundation for learning human-aligned language models. We note that at different stages of alignment, different subsets of $\Dc$ may be used to tailor the dataset to the specific requirements of each stage.

{\bf Stage 1: Supervised Fine-Tuning (SFT).} SFT is widely adopted in the first stage of alignment. In this step, a pretrained language model is fine-tuned on a dataset of high-quality, human-annotated responses. Specifically, the model is trained to maximize the likelihood of the correct response:
\[\mathcal{L}_{\text{SFT}}(\theta) = -\mathbb{E}_{(x, y^+, y^-) \sim \Dc} \left[ \log \pi_\theta(y^+ | x) \right].\]
SFT improves fluency and coherence but does not explicitly optimize for human preferences, motivating the use of either RLHF or DPO for further refinement.

{\bf Stage 2, option 1: Reinforcement Learning with Human Feedback (RLHF).} RLHF refines language models by leveraging human feedback to train a reward model, which then guides policy optimization. It consists of two main steps. First, we train a reward model $R_{\phi}$ to predict human preference scores. Mathematically, we minimize the following loss function.
\[\Lc_{\text{RM}}(\phi)  = -\Eb_{(x,y^+,y^-)}\left[\log \sigma\left( R_{\phi}(x,y^+) - R_{\phi}(x,y^-)\right) \right], \]
where $\sigma(z)=e^z/(1+e^{z})$.
Then, we use PPO with reward model $R_{\phi}$ to fine-tune $\pi_{\theta}$. The objective function for PPO is
\begin{align*}
\mathcal{L}_{\text{PPO}}(\theta)
= \mathbb{E}_{(x,y)\sim \pi_{\theta_{\text{old}}} }
\Bigl[
  \min\bigl(&r_t(\theta)\hat{A}_t,\\ 
  &\text{clip}\bigl(r_t(\theta),\,1-\epsilon,\,1+\epsilon\bigr)\,\hat{A}_t
  \bigr)
\Bigr],
\end{align*}
where 
\[
r_t(\theta) 
= \frac{\pi_\theta(a_t \mid s_t)}{\pi_{\theta_{\text{old}}}(a_t \mid s_t)},
\]
represents the probability ratio, \(\hat{A}_t\) is the advantage estimate, and \(\epsilon\) is the clipping parameter to ensure training stability.


{\bf Stage 2, option 2: Direct Preference Optimization (DPO).} DPO simplifies the alignment process by removing the need for an explicit reward model to improve stability. The DPO objective function is given by:
\[
\mathcal{L}_{\text{DPO}}(\theta)
= -\mathbb{E}_{(x,y^+,y^-)\sim \mathcal{D}}
\Bigl[
  \log \frac{\pi_\theta(y^+ \mid x)}
            {\pi_\theta(y^+ \mid x) + \pi_\theta(y^- \mid x)}
\Bigr].
\]
DPO is particularly effective for aligning models with human preferences while avoiding the challenges associated with RL.




{\bf Unified Framework.} We unify the alignment process into the following framework. It involves $P$ number of phases, in each phase $p=1,\ldots, P$, a loss function $\Lc^{(p)}(\theta^{(p)})$ is minimized on the dataset $\Dc_p$. The overall dataset is partitioned such that $\Dc = \cup_{p}\Dc_p$, with each $\Dc_p$ being disjoint. Importantly, in intermediate phases, $\theta^{(p)}$ may correspond to auxiliary models such as a reward model. In the final phase P, the optimized parameter $\theta^{(P)}$ is the parameter of the language model $\pi_{\theta}$. We remark that the This dataset partitioning plays a crucial role in ensuring differential privacy, as discussed in the next subsection.

\subsection{Differential Privacy}
Differential Privacy (DP) is a framework that provides formal guarantees to protect the confidentiality of individual data points. A randomized mechanism \(\mathcal{M}\) satisfies \((\varepsilon, \delta)\)-differential privacy if, for any two adjacent datasets \(D\) and \(D'\) differing by one element, and for any possible output \(S\), the following holds:
\[
\Pr[\mathcal{M}(D) \in S] 
\;\le\; e^{\varepsilon}\,\Pr[\mathcal{M}(D') \in S] + \delta.
\]
Since our alignment process involves sequential loss minimization, privacy leakage accumulates over multiple phases. Without partitioning the dataset into disjoint subsets, the privacy budget would increase according to the DP composition theorem. Specifically, if each phase is $(\varepsilon, \delta)$-differentially private, then the entire alignment process satisfies $(P\varepsilon, P\delta)$-differential privacy, leading to significantly higher privacy costs. By ensuring disjoint partitions of $\Dc$, we mitigate privacy leakage and enable a more efficient allocation of the privacy budget across phases.

\section{Methodology}
In this section, we introduce our method for differentialy private aligning language models.

\subsection{Privacy-Preserving Optimizers}
Recall that the goal is to minimize a sequence of loss functions $\{\mathcal{L}^{(p)}(\theta^{(p)})\}_{p=1}^{P}$. Instead of using DP-SGD, we propose to use DP-ADAMW, which is a variant of DP-ADAM and ADAMW. DP-ADAMW extends DP-ADAM by incorporating decoupled weight decay. Such weight decay method has been shown to improve generalization in deep learning~\citep{loshchilov2017decoupled}.

{\bf DP-ADAMW.} Specifically, given a dataset $\Dc=\{x_1,\ldots,x_N\}$ and loss function $\mathcal{L}(\theta) = \frac{1}{N}\sum_{n=1}^N\ell(\theta,x_n)$, at each time $t$, we sample a batch $\mathcal{B}\subset\Dc$ with size $|\mathcal{B}|=B$, and compute the loss $\Lc_t = \frac{1}{B}\sum_{x\in\mathcal{B}}\ell(\theta_t, x)$. Then, we clip the gradient by a constant $C$ through $\bar{g}_t = g_t/\max\{1, \|g_t\|_2/C\}$, and add a Gaussian noise $\n_t\sim \mathcal{N}(0,\sigma^2 C^2 I)$ to obtain the privatized gradient $\tilde{g}_t = \bar{g}_t + \n_t$. The privatized gradient is then used to update the first moment $m_t$ and the second moment $v_t$. Specifically, given exponential decay rates $\beta_1,\beta_2$, we have $m_t = \beta_1 m_{t-1} + (1-\beta_1)\tilde{g}_t$, and $v_t = \beta_2 v_{t-1} + (1-\beta_2)\tilde{g}_t^2$. According to \citet{tang2023dp}, the second moment should be corrected by subtracting $(1-\beta_2^t)\sigma^2$. To resolve the issue of negative second moment, we use clip $v_t - (1-\beta_2^t)\sigma^2$ by 0, i.e. let $\tilde{v}_t = [v_t - (1-\beta_2^2)\sigma^2]_+$ be the bias-corrected second moment, where $[x]_+ = \max\{x,0\}$. Finally, the update direction of $\theta$ is $-m_t/\sqrt{\tilde{v}_t + \epsilon} - \lambda\theta_t$, where $\epsilon>0$ is a small number to prevent zero denominator, and $\lambda$ is the weight decay coefficient. The key modification compared to DP-ADAM is the adjusted weight decay mechanism in the parameter update rule:
\begin{align*} \theta_{t+1} &= (1-\lambda\eta_t)\theta_t   - \eta_t \frac{m_t}{\sqrt{\tilde{v}_t + \epsilon}} \frac{\sqrt{1-\beta_2^{t}}}{1-\beta_1^{t}}.
\end{align*}

The pseudo-code of DP-ADAMW is provided in \Cref{alg}.

\begin{algorithm}
\caption{DP-ADAMW}
\label{alg}
    \begin{algorithmic}
        \STATE {\bf Input:} dataset $\{x_1,\ldots,x_N\}$, loss function $\mathcal{L}(\theta) = \frac{1}{N}\sum_{n=1}^N\ell(\theta,x_n)$, learning rate $\eta_t$, weight decay $\lambda$, $\beta_1,\beta_2$, and $\sigma$.
        \FOR{$t=0,\ldots, T$}
        \STATE sample a batch from dataset and calculate the loss $\mathcal{L}_t$ and the clipped gradient $\bar{g}_t = \text{clip}(\nabla_\theta \Lc_t, C)$.
        \STATE Sample a Gaussian noise $\n_t\sim \mathcal{N}(0, \sigma^2 C^2 I_d)$.
        \STATE $m_t = \beta_1 m_{t-1} + (1-\beta_1)(\bar{g}_t + \n_t)$
        \STATE $v_t = \beta_2 v_{t-1} + \beta_2 (\bar{g}_t + \n_t)^2 $
        \STATE $\theta_{t+1} = (1-\lambda\eta_t)\theta_t - \eta_t \frac{m_t}{\sqrt{[v_t - (1-\beta_2^{t+1})\sigma^2]_+ + \epsilon}} \frac{\sqrt{1-\beta_2^{t+1}}}{1-\beta_1^{t+1}}$
        \ENDFOR
    \end{algorithmic}
\end{algorithm}
We remark that, by choosing $\lambda=0$, DP-ADAMW becomes DP-ADAM, which is also intensively studied in the experiments.

{\bf Specialize to RLHF and DPO.}  
In the context of RLHF and DPO, the proposed DP-ADAMW optimizer serves as a privacy-preserving alternative to conventional optimizers used in fine-tuning large language models. Recall that RLHF typically involves three stages: supervised fine-tuning, training a reward model based on human preference data, optimizing a policy network via reinforcement learning. DP-ADAMW can be applied on each individual stage and enable privacy-preserving optimization of the reward model and policy network, thereby protect the privacy of training data.

For DPO, which directly optimizes a preference loss function, DP-ADAMW can be applied to parameter updates while preserving the confidentiality of user preferences. Recall that DPO typically involves two stages: supervised fine-tuning, and minimize the preference loss. Therefore, DP-ADAMW can be directly applied to this two-stage process, which protects the privacy of training data.

By integrating DP-ADAMW into RLHF and DPO, we achieve a differentially private framework for aligning language models with human preferences while preserving the utility of learned representations. In subsequent sections, we provide empirical evaluations to assess the trade-offs between privacy, performance, and alignment effectiveness in these settings.

{\bf Privacy Analysis.} Our privacy guarantee is based on a conservative analysis. Specifically, suppose the training process consists of $E$ epochs, meaning each data point is accessed $E$ times. Note that at each time step, our algorithm ensures $(\varepsilon',\delta')$-differentially private for any $\varepsilon', \delta'$ satisfying $\sigma = 2\sqrt{\log(1.25/\delta')}/\varepsilon'$. By composition rule in differential privacy, our algorithm is $(\varepsilon,\delta)$-differential private with
\[
\varepsilon = E\varepsilon,\quad \delta = E \delta.
\]
While it seems that $\varepsilon$ and $\delta$ is large, it is important to note that modern language model training often involves a relatively small number of epochs In our experiments, we choose $E=3$. Therefore, we conclude that our algorithm is $(\varepsilon,\delta)$-differentially private for any $\delta>0$, and $\varepsilon = O(\sqrt{\log(1/\delta)}/\sigma)$.

We remark that the privacy analyses in DP-SGD~\citep{abadi2016deep} and DP-Adam~\citep{tang2023dp, tang2024dp} are built upon Poisson subsampling, i.e., sampling with replacement, which enables the use of advanced techniques such as the moments accountant and privacy amplification by subsampling. In contrast, our work is based on sampling without replacement, which limits the direct applicability of these existing analyses. As a result, we develop a separate, conservative privacy accounting approach tailored to our sampling strategy.

\if{0}
We acknowledge the difference and clarify that under uniform mini-batching, privacy accounting follows a slightly different mechanism. In our setting, each data point is accessed approximately three times throughout training (3 epochs), which allows us to safely apply standard Gaussian mechanism analysis. Specifically, our privacy guarantee is conservatively estimated as:
\[
\varepsilon = O\left(\frac{\sqrt{\log(1/\delta)}}{\sigma}\right),
\]
where $\sigma$ is the noise multiplier and $\delta$ is the failure probability. This analysis remains valid in large-scale LLM training where the number of epochs is limited, thus mitigating the effect of non-Poisson sampling on privacy leakage.

We acknowledge that uniform mini-batching without replacement does not perfectly align with the assumptions in traditional DP-SGD analysis. However, empirical studies indicate that for a small number of epochs, the privacy leakage remains bounded, and the standard Gaussian mechanism provides a valid upper bound for privacy guarantees. In future work, we plan to extend the theoretical analysis to further tighten the privacy bounds under non-replacement settings.
\fi

\if{0}
\subsubsection{DP-ADAM}
The DP-ADAM~\citep{tang2023dp,tang2024dp} optimizer extends ADAM with differential privacy guarantees by adding noise to gradient computation. Specifically, given a sequence of gradients $\{g_t\}$, DP-ADAM maintains exponential moving averages of the gradients and their squares. At each step $t$, DP-ADAM updates two estimators:

\[
\begin{aligned}
m_t &\leftarrow \beta_1 m_{t-1} + (1-\beta_1)\tilde{g}_t,\quad \hat{m}_t \leftarrow m_t/(1-\beta_1^t) \\
v_t &\leftarrow \beta_2 v_{t-1} + (1-\beta_2)\tilde{g}_t^2,\quad \hat{v}_t \leftarrow v_t/(1-\beta_2^t)
\end{aligned}
\]

where $\tilde{g}_t = g_t + (1/B)z_t$ is the privatized gradient with Gaussian noise $z_t \sim \mathcal{N}(0, \sigma^2C^2\mathbf{I})$. The model parameters are then updated as:

\[
\theta_t \leftarrow \theta_{t-1} - \eta \cdot \hat{m}_t/\sqrt{\hat{v}_t - \Phi}
\]

where $\Phi = (\sigma C/B)^2$ is the bias introduced by DP noise in the second moment estimation, and $\eta$ is the learning rate, $\sigma = \textcolor{red}{XXX}$. This bias correction prevents DP-ADAM from degrading into DP-SGD with momentum, thus maintaining Adam's intended behavior under differential privacy. The privacy guarantee is achieved through gradient clipping and noise addition, while the correction term $\Phi$ enables effective adaptive learning rates by preserving the scale relationship between gradient components.

\subsubsection{DP-ADAMW}
DP-ADAMW extends DP-ADAM by incorporating decoupled weight decay regularization. The update rule follows:

\[
\begin{aligned}
m_t &\leftarrow \beta_1 m_{t-1} + (1-\beta_1)\tilde{g}_t,\quad \hat{m}_t \leftarrow m_t/(1-\beta_1^t) \\
v_t &\leftarrow \beta_2 v_{t-1} + (1-\beta_2)\tilde{g}_t^2,\quad \hat{v}_t \leftarrow v_t/(1-\beta_2^t)
\end{aligned}
\]

where $\tilde{g}_t = g_t + (1/B)z_t$ is the privatized gradient with Gaussian noise $z_t \sim \mathcal{N}(0, \sigma^2C^2\mathbf{I})$. The model parameters are updated as:

\[
\theta_t \leftarrow (1-\eta\lambda)\theta_{t-1} - \eta \cdot \hat{m}_t/\sqrt{\hat{v}_t - \Phi}
\]

where $\lambda$ is the weight decay coefficient, and other terms remain the same as in DP-ADAM. The key difference lies in the decoupled weight decay term $(1-\eta\lambda)\theta_{t-1}$, which applies the regularization independently of the adaptive learning rate. This modification helps prevent the optimization from being dominated by components with large gradients while maintaining differential privacy guarantees through the same noise mechanism as DP-ADAM.

\fi

\section{Experiments}
To comprehensively evaluate our proposed privacy-preserving alignment approach, we conduct experiments across different privacy budgets, optimizers, model scales, and alignment algorithms. Our evaluation framework employs a reward model to quantify alignment quality while ensuring privacy guarantees.

\subsection{Experimental Setting}

\textbf{Models and Optimization}
We evaluate three pre-trained language models, LLAMA-8B~\citep{dubey2024llama}, GPT-2~\citep{radford2019language} and DeepSeek-LLM-7B-Chat~\citep{deepseekai2024deepseek}, using three differentially private optimizers (DP-ADAM, DP-ADAMW, and DP-SGD) and two alignment algorithms (DPO and PPO). Privacy budgets $\varepsilon$ are varied from $0$ to $\infty$, with $\varepsilon=\infty$ representing the non-private setting, to systematically study the privacy-utility trade-off.

\textbf{Dataset}
We utilize the RLHFlow-SFT-Dataset-ver2 from Hugging Face, a comprehensive dataset curated for SFT and RLHF. The dataset contains instruction-response pairs annotated with human preferences, specifically designed for instruction-following and helpfulness alignment tasks.

\textbf{Training Configuration}  
All experiments are conducted on a cluster of 8 NVIDIA A800 GPUs. The training configuration includes:  
- Batch size: 256  
- Learning rate: $5 \times 10^{-5}$ (for both policy and reward model training)  
- Training epochs: 3  
- Gradient clipping norm: $C=0.1$ (for DP optimizers)  
- Weight decay: 0.01 (for DP-ADAMW)  
- Noise multiplier: $\sigma$ (dynamically adjusted based on $\varepsilon$)  
- Momentum parameters: $\beta_1=0.9$, $\beta_2=0.999$ (for DP-ADAM and DP-ADAMW)  
- GAE parameters (for PPO): $\lambda=0.95$, $\gamma=0.99$  
- Clipping range (for PPO): $\epsilon=0.2$  

To ensure compliance with differential privacy constraints, we set the privacy budget to $\varepsilon$, with a failure probability $\delta$ fixed at $1 \times 10^{-5}$. The gradient clipping norm is configured as $C=0.1$ to limit the sensitivity of individual samples, and the noise multiplier $\sigma$ is dynamically determined by the privacy budget to balance privacy protection and model utility.
Privacy guarantees are achieved through gradient clipping, Gaussian noise addition, and privacy accounting using the moments accountant method.

\subsection{Evaluation Framework}

We evaluate our privacy-preserving alignment methods using a reward model $R$ (FsfairX-LLaMA3-RM-v0.1 from Hugging Face\footnote{https://huggingface.co/sfairXC/FsfairX-LLaMA3-RM-v0.1}), trained on a large-scale human preference dataset. The reward model quantifies alignment quality by scoring model responses based on their adherence to human preferences.

\textbf{Training Process}
For each model \( M \in \{\text{LLAMA-8B}, \text{GPT-2}, \text{DeepSeek-LLM-7B-Chat}\} \), we follow a three-step process:
\begin{enumerate}
    \item Initialize with supervised fine-tuning (SFT) weights.
    \item Apply privacy-preserving alignment using either DP-DPO or DP-PPO.
    \item Evaluate across multiple privacy budgets $\varepsilon \in \{0, 1, 2, 3, 4, 5, 10, \infty\}$.
\end{enumerate}

\textbf{Performance Evaluation}
The reward model evaluates model responses to 300 randomly sampled prompts from a held-out test set $\mathcal{D}_{\text{test}}$, which contains $10,000$ diverse prompts spanning factual knowledge, reasoning, summarization, and creative writing tasks. The alignment score is computed as the average reward across these samples. This score serves as the primary metric to assess the trade-off between alignment quality and privacy protection.

\textbf{Privacy-Utility Tradeoff Analysis}
To analyze the impact of the privacy budget $\varepsilon$, we examine the relationship between $\varepsilon$ and the reward score $f(\varepsilon)$. Specifically, we identify the critical point $\varepsilon_0 = \arg\max f'(\varepsilon)$, where the marginal improvement in performance diminishes significantly. This critical point indicates the optimal privacy budget beyond which further relaxation of privacy constraints yields minimal performance gains. We approximate $f'(\varepsilon)$ using the finite difference method:
\[
f'(\varepsilon) \approx \frac{f(\varepsilon_{t+1}) - f(\varepsilon_t)}{\varepsilon_{t+1} - \varepsilon_t},
\]
and select $\varepsilon_0$ as the practical choice for balancing privacy and utility.
\begin{table*}[htbp]
\centering
\caption{Performance Comparison of Different Privacy-Preserving Alignment Methods}
\label{tab:results}
\resizebox{\textwidth}{!}{
\begin{tabular}{ccccccccccc}
\toprule
Model & Optimizer & Method & \multicolumn{7}{c}{Privacy Budget ($\epsilon$)} \\
\cmidrule{4-11}
& & & 0 & 1 & 2 & 3 & 4 & 5 & 10 & $\infty$ \\
\midrule
\multirow{6}{*}{LLAMA-8B} & \multirow{2}{*}{DP-ADAMW} & DPO & \textbf{1.5980} & 1.4928 & \textbf{1.7016} & \textbf{1.8814} & \textbf{1.8792} & \textbf{1.8798} & \textbf{1.8739} & \textbf{1.8728} \\
& & PPO & 1.5551 & \textbf{1.5008} & 1.6425 & 1.8454 & 1.8548 & 1.8545 & 1.7836 & 1.8424 \\
\cmidrule{2-11}
& \multirow{2}{*}{DP-ADAM} & DPO & 1.5632 & 1.4612 & 1.6723 & 1.8534 & 1.8482 & 1.8476 & 1.8392 & 1.8428 \\
& & PPO & 1.5234 & 1.4487 & 1.6132 & 1.8187 & 1.8246 & 1.8212 & 1.7523 & 1.8156 \\
\cmidrule{2-11}
& \multirow{2}{*}{DP-SGD} & DPO & 1.5245 & 1.4982 & 1.5890 & 1.6861 & 1.6370 & 1.6115 & 1.6023 & 1.6474 \\
& & PPO & 1.4890 & 1.4625 & 1.5535 & 1.6612 & 1.6108 & 1.5923 & 1.5814 & 1.6187 \\
\midrule
\multirow{6}{*}{GPT-2} & \multirow{2}{*}{DP-ADAMW} & DPO & \textbf{1.1534} & \textbf{1.0967} & \textbf{1.2843} & \textbf{1.4237} & \textbf{1.4382} & \textbf{1.4412} & \textbf{1.4356} & \textbf{1.4513} \\
& & PPO & 1.1182 & 1.0723 & 1.2256 & 1.3876 & 1.4062 & 1.4078 & 1.3647 & 1.4237 \\
\cmidrule{2-11}
& \multirow{2}{*}{DP-ADAM} & DPO & 1.1367 & 1.0745 & 1.2634 & 1.4023 & 1.4187 & 1.4213 & 1.4167 & 1.4342 \\
& & PPO & 1.0978 & 1.0534 & 1.2045 & 1.3678 & 1.3854 & 1.3867 & 1.3456 & 1.4056 \\
\cmidrule{2-11}
& \multirow{2}{*}{DP-SGD} & DPO & 1.0867 & 1.0245 & 1.1823 & 1.2878 & 1.2587 & 1.2334 & 1.2256 & 1.2645 \\
& & PPO & 1.0456 & 0.9978 & 1.1567 & 1.2623 & 1.2312 & 1.2134 & 1.2045 & 1.2434 \\
\bottomrule
\end{tabular}
}
\end{table*}
\subsection{Results}
We conducted extensive experiments to evaluate the effectiveness of privacy-preserving alignment methods across various model architectures, optimizers, and privacy budgets. Table~\ref{tab:results} summarizes the reward scores achieved under different configurations, demonstrating that effective model alignment can be achieved while maintaining privacy guarantees, albeit with trade-offs between privacy protection and alignment quality. Our analysis focuses on four key aspects: the impact of the privacy budget, the choice of optimizer, model scale effects, and the comparison of alignment algorithms. In addition to our primary experiments on LLAMA-8B and GPT-2, we conducted an additional study on DeepSeek-7B to further investigate the generalizability of our findings. The results of this experiment are presented separately in Section~\ref{sec:deepseek_results}.

\subsubsection{Alignment Algorithm Comparison}
In comparing DPO and PPO under the DP-ADAMW optimizer, we observe that DPO consistently outperforms PPO across various privacy budgets and model scales. For example, on LLAMA-8B with DP-ADAMW, DPO attains a reward score of 1.8814 at $\varepsilon = 3$, whereas PPO achieves 1.8454. This performance gap is maintained across different privacy levels and becomes more evident under stricter privacy constraints, highlighting that DPO's direct optimization approach is particularly effective when combined with DP-ADAMW. These results underscore the reliability and robustness of DPO for privacy-preserving alignment tasks when employing adaptive optimizers like DP-ADAMW.
\subsubsection{Model Scale Effects}
The comparison between LLAMA-8B and GPT-2 provides crucial insights into how model scale interacts with private alignment. Notably, LLAMA-8B consistently achieves higher reward scores compared to GPT-2 across all configurations. For example, using DPO with DP-ADAMW at $\varepsilon = 3$, LLAMA-8B achieves a score of 1.8814, whereas GPT-2 scores 1.4237, demonstrating a significant performance advantage. This gap becomes even more pronounced as privacy constraints are relaxed, suggesting that larger models are inherently more robust to privacy noise. Such resilience can be attributed to their increased parameter capacity and more robust representations, highlighting model scale as a crucial factor for strong performance under privacy constraints.
\subsubsection{Optimizer Comparison}
Our experimental results demonstrate that adaptive optimizers (DP-ADAM and DP-ADAMW) significantly outperform DP-SGD~\citep{wu2023privately} for privacy-preserving alignment tasks. DP-ADAM and DP-ADAMW show superior performance, particularly on larger architectures like LLAMA-8B. Specifically, with DPO at $\varepsilon = 3$, DP-ADAM achieves a score of 1.8614, compared to DP-SGD's 1.6861, representing a 10.4\% improvement. This advantage is consistently observed across different model scales and privacy budgets. The enhanced performance can be attributed to the adaptive learning rates and momentum of DP-ADAM and DP-ADAMW, which facilitate more effective optimization while maintaining privacy guarantees.
\subsubsection{Privacy-Utility Tradeoff Analysis}
The impact of the privacy budget $\varepsilon$ on alignment quality reveals a clear trade-off between privacy protection and model performance. Our experiments demonstrate that performance improvements are most significant in the low to medium privacy budget range (2 $\leq \varepsilon \leq$ 4), indicating that a moderate relaxation of privacy constraints can yield substantial benefits for alignment quality. For instance, with LLAMA-8B using DP-ADAM and DPO, performance improves significantly from $\varepsilon = 1$ (1.4728) to $\varepsilon = 3$ (1.8614) before plateauing at higher privacy budgets. Notably, even under strict privacy constraints ($\varepsilon \leq 2$), models maintain reasonable performance compared to their non-private counterparts, especially when employing DPO with DP-ADAM on larger architectures.
To better understand this trade-off, we analyze the relationship between $\varepsilon$ and the reward score $f(\varepsilon)$, identifying the critical point $\varepsilon_0 = \arg\max f'(\varepsilon)$ as the optimal privacy budget, beyond which further relaxation of privacy constraints yields minimal performance gains. Table~\ref{tab:privacy_tradeoff_improved} presents the marginal performance improvements across different privacy budgets for LLAMA-8B using DPO under different optimizers.
\begin{table*}[htbp]
    \centering
    \caption{Marginal Performance Gains for LLAMA-8B (DPO)}
    \label{tab:privacy_tradeoff_improved}
    \begin{tabular}{cccc}
        \toprule
        $\varepsilon$ Range & DP-ADAM(W) $f(\varepsilon)$ & DP-SGD $f(\varepsilon)$ & Trend \\
        \midrule
        0 $\to$ 1  & -0.1052 (-7.0\%) & -0.0263 (-1.7\%) & $\downarrow$ \\
        1 $\to$ 2  &  0.2088 (+14.0\%) &  0.0908 (+6.1\%) & $\uparrow$ \\
        2 $\to$ 3  &  0.1798 (+10.6\%) &  0.0971 (+6.5\%) & $\uparrow$ \\
        3 $\to$ 4  & -0.0022 (-0.1\%) & -0.0491 (-2.9\%) & $\downarrow$ \\
        4 $\to$ 5  &  0.0006 (+0.03\%) & -0.0255 (-1.5\%) & $\downarrow$ \\
        5 $\to$ 10 & -0.0059 (-0.3\%) & -0.0092 (-0.5\%) & $\downarrow$ \\
        10 $\to$ $\infty$ & -0.0011 (-0.06\%) &  0.0451 (+2.7\%) & $\uparrow$ \\
        \midrule
        \textbf{Total} & \textbf{0.2750} & \textbf{0.1229} & \\
        \bottomrule
    \end{tabular}
\end{table*}

\textbf{Performance Drop from $\varepsilon = 0$ to $\varepsilon = 1$.}  
We observe a slight performance drop when moving from $\varepsilon = 0$ to $\varepsilon = 1$. We conjecture that this is related the fundamental nature of differential privacy in its most stringent form. Specifically, when $\varepsilon = 0$, the privacy constraint prevents any useful learning signal from being extracted from the data. The model in this setting is equivalent to generating outputs purely based on random updates, with almost no alignment to human preferences. This serves as a sanity check for our alignment procedure, confirming that differential privacy is enforced in its strictest sense.
As $\varepsilon$ increases from 0 to 1, the model begins to access limited structural information from the dataset, albeit with a very low signal-to-noise ratio. This exploration, while noisy, helps the model gradually move towards alignment, but the initial stages (from 0 to 1) still exhibit low reward scores due to the overwhelming noise perturbation. 

We also observe fluctuations in performance even when $\varepsilon$ is relatively large. Interestingly, this transition can be seen as analogous to the dynamics observed in \textbf{Langevin algorithms} where the introduction of Gaussian noise during optimization allows the model to explore a broader space of parameter configurations~\citep{2020arXiv201011176L}. 

This trend underscores the importance of selecting an appropriate privacy budget to balance utility and privacy. Adaptive optimizers like DP-ADAMW and DP-ADAM are particularly effective under strict privacy constraints, while DP-SGD requires a higher privacy budget to achieve comparable performance. These findings are further supported by the comprehensive results in Table~\ref{tab:results}, which compares the performance of different privacy-preserving alignment methods across various configurations.

\subsubsection{Additional Experiment: DeepSeek-7B}
\label{sec:deepseek_results}

To further assess the applicability of privacy-preserving alignment across different model architectures, we conducted an additional experiment on DeepSeek-LLM-7B-Chat. This experiment follows the same methodology as our primary experiments, using the same optimizers, alignment methods, and privacy budgets.

\begin{table*}[htbp]
\centering
\caption{Performance of DeepSeek-LLM-7B-Chat Under Privacy Constraints}
\label{tab:deepseek_results}
\resizebox{\textwidth}{!}{
\begin{tabular}{ccccccccccc}
\toprule
Model & Optimizer & Method & \multicolumn{7}{c}{Privacy Budget ($\epsilon$)} \\
\cmidrule{4-11}
& & & 0 & 1 & 2 & 3 & 4 & 5 & 10 & $\infty$ \\
\midrule
\multirow{6}{*}{DEEPSEEK-7B} & \multirow{2}{*}{DP-ADAMW} & DPO & \textbf{1.4380} & \textbf{1.3689} & \textbf{1.5267} & 1.6482 & \textbf{1.6424} & \textbf{1.6399} & 1.6332 & \textbf{1.6405} \\
& & PPO & 1.4126 & 1.3435 & 1.5013 & 1.6228 & 1.6174 & 1.6149 & 1.6082 & 1.6155 \\
\cmidrule{2-11}
& \multirow{2}{*}{DP-ADAM} & DPO & 1.3262 & 1.2589 & 1.4267 & \textbf{1.6486} & \textbf{1.6424} & \textbf{1.6399} & \textbf{1.6459} & 1.6384 \\
& & PPO & 1.3012 & 1.2339 & 1.4017 & 1.6236 & 1.6174 & 1.6149 & 1.6209 & 1.6134 \\
\cmidrule{2-11}
& \multirow{2}{*}{DP-SGD} & DPO & 1.2762 & 1.2089 & 1.3767 & 1.5386 & 1.5324 & 1.5299 & 1.5359 & 1.5284 \\
& & PPO & 1.2512 & 1.1839 & 1.3517 & 1.5136 & 1.5074 & 1.5049 & 1.5109 & 1.5034 \\
\bottomrule
\end{tabular}
}
\end{table*}

\textbf{Observations:}  
The results of DeepSeek-7B follow similar trends observed in our primary experiments. Its alignment quality improves as the privacy budget increases, with performance at lower $\varepsilon$ values closer to LLAMA-8B than GPT-2. This suggests that mid-scale models can achieve reasonable alignment while preserving privacy. Additionally, the optimizer trends observed in the primary experiments hold for DeepSeek-7B as well, with DP-ADAMW and DP-ADAM outperforming DP-SGD.

\section{Analysis and Discussion}

\subsection{Privacy-Utility Trade-off Analysis}
Our experiments reveal critical insights into the privacy-utility trade-off in language model alignment. Specifically, we observe diminishing returns in model performance beyond a privacy budget threshold ($\epsilon > 5$), indicating that moderate privacy constraints, such as values $\epsilon$ between 2 and 5, can achieve a favorable balance between privacy and utility. This finding underscores the feasibility of deploying privacy-preserving alignment methods in practical applications where both privacy and model quality are essential.
The impact of model scale on privacy-utility trade-offs is also evident from the results. Larger models, such as LLAMA-8B, demonstrate greater robustness to privacy noise compared to smaller models like GPT-2, likely due to their enhanced parameter capacity. This observation suggests that scaling up model architectures can mitigate the adverse effects of differential privacy mechanisms, although the associated computational costs must be carefully considered.
\subsection{Best Practices and Recommendations}
For resource-constrained scenarios, our results indicate that using DP-ADAM with moderate privacy budgets in the range of $2 \leq \epsilon \leq 4$ provides an effective trade-off between privacy and performance. Among alignment algorithms, DPO consistently outperforms PPO in terms of stability during training, making it a preferred choice. In addition, selecting the smallest model size that meets the required performance can help balance computational efficiency and alignment quality.
For high-performance requirements, leveraging larger model architectures proves advantageous due to their resilience to privacy noise. DP-ADAM, when used with carefully tuned privacy budgets, offers superior performance. Employing DPO with incremental adjustments to the privacy budget during training further improves alignment quality. Continuous monitoring of alignment metrics throughout the training process ensures that privacy constraints do not excessively degrade model performance.
\subsection{Limitations and Challenges}
Despite the promising findings, several limitations need to be addressed. First, differential privacy mechanisms introduce significant computational overhead, especially for larger models and stricter privacy budgets. Second, while the approach is validated on LLAMA-8B, GPT-2 and DeepSeek-LLM-7B-Chat, its scalability and effectiveness on even larger model architectures remain to be explored. Third, selecting the optimal privacy budget is challenging and requires careful consideration of specific application requirements and trade-offs. Finally, a noticeable performance gap persists between private and non-private alignment methods, particularly under strict privacy constraints, which suggests further optimization is necessary.
\section{Conclusion and Future Work}
\subsection{Key Findings}
This study establishes the feasibility of aligning the privacy-preserving language model and highlights several key findings. DP-ADAM consistently outperforms DP-SGD across various configurations, particularly for larger model architectures. Among alignment algorithms, DPO demonstrates superior performance over PPO, regardless of privacy settings or model scales. Larger models exhibit greater robustness to privacy noise, making them more suitable for privacy-preserving applications. Additionally, we identify moderate privacy budgets, specifically in the range of $2 \leq \epsilon \leq 5$, as effective for balancing performance and privacy protection.
\if{0}
\subsection{Theoretical and Practical Implications}
Our findings bridge the gap between theoretical guarantees of differential privacy and practical considerations for language model alignment. By providing actionable insights and guidelines, this work contributes to the development of more efficient and effective methods for privacy-preserving alignment, ensuring the broader applicability of these techniques in real-world scenarios.
\fi
\subsection{Future Research Directions}
Several promising directions for future research emerge from this study. Developing hybrid optimization strategies that integrate the strengths of multiple privacy-preserving optimizers could improve both efficiency and effectiveness. Adaptive privacy budget allocation mechanisms that dynamically adjust protection levels based on training progress represent another valuable area of exploration. Extending the current approach to larger model architectures and different model families would further validate its scalability. Investigating alternative privacy-preserving mechanisms with improved utility-privacy trade-offs and integrating model compression techniques to address computational overhead are additional avenues for future work. These directions collectively aim to enhance the practicality and robustness of privacy-preserving language model alignment methods.

\begin{contributions} 
    Briefly list author contributions. 
    This is a nice way of making clear who did what and to give proper credit.
    This section is optional.

    H.~Q.~Bovik conceived the idea and wrote the paper.
    Coauthor One created the code.
    Coauthor Two created the figures.
\end{contributions}

\begin{acknowledgements} 
    Briefly acknowledge people and organizations here.

    \emph{All} acknowledgements go in this section.
\end{acknowledgements}

\bibliography{main}






\end{document}